\documentclass[final]{cvpr}

\usepackage{times}
\usepackage{epsfig}
\usepackage{graphicx}
\usepackage{gensymb}
\usepackage{amsmath,amssymb} 
\usepackage{graphicx}
\usepackage[utf8]{inputenc} 
\usepackage[T1]{fontenc}    
\usepackage{comment}
\usepackage{amsfonts}       
\usepackage{nicefrac}       
\usepackage{microtype}      
\usepackage{url}            
\usepackage{booktabs}       
\usepackage{amsfonts}       
\usepackage{nicefrac}       
\def\ie{\emph{i.e.}}
\usepackage{multirow}
\usepackage{xcolor}
\usepackage{subfig}

\usepackage[misc]{ifsym}
\usepackage{enumitem}

\usepackage[pagebackref=true,breaklinks=true,colorlinks,bookmarks=false]{hyperref}

\def\cvprPaperID{845} 
\def\confYear{CVPR 2021}
\begin{document}

\title{Visually Informed Binaural Audio Generation without Binaural Audios}  

\author{Xudong Xu$^{1}$\thanks{Equal contribution.} \quad Hang Zhou$^{1}$\footnotemark[1] \quad Ziwei Liu$^2$ \quad  Bo Dai$^{2}$ \quad Xiaogang Wang$^{1}$ \quad Dahua Lin$^{1}$\\
    $^1$CUHK - SenseTime Joint Lab, The Chinese University of Hong Kong \\
    $^2$S-Lab, Nanyang Technological University\\
    {\tt\small \{xx018@ie,zhouhang@link,xgwang@ee,dhlin@ie\}.cuhk.edu.hk},\hspace{5pt}
    {\tt\small \{ziwei.liu,bo.dai\}@ntu.edu.sg}
}
\maketitle

\begin{abstract}

Stereophonic audio, especially binaural audio, plays an essential role in immersive viewing environments. 
%
Recent research has explored generating visually guided stereophonic audios supervised by multi-channel audio collections.
However, due to the requirement of professional recording devices, 
existing datasets are limited in scale and variety, which impedes the generalization of supervised methods in real-world scenarios.
%
In this work, we propose \textbf{PseudoBinaural}, an effective pipeline that is free of binaural recordings.
The key insight is to carefully build pseudo visual-stereo pairs with mono data for training.
Specifically, we leverage spherical harmonic decomposition and head-related impulse response (HRIR) to identify the relationship between spatial locations and received binaural audios.
%
Then in the visual modality, corresponding visual cues of the mono data are manually placed at sound source positions to form the pairs.
Compared to fully-supervised paradigms, our binaural-recording-free pipeline shows great stability in cross-dataset evaluation and achieves comparable performance under subjective preference. 
Moreover, combined with binaural recordings, our method is able to further boost the performance of binaural audio generation under supervised settings\footnote{Code, models and demo videos are available at \url{https://sheldontsui.github.io/projects/PseudoBinaural}.}.

\end{abstract}

\begin{figure*}[t]
    \centering
    \includegraphics[width=0.95\linewidth]{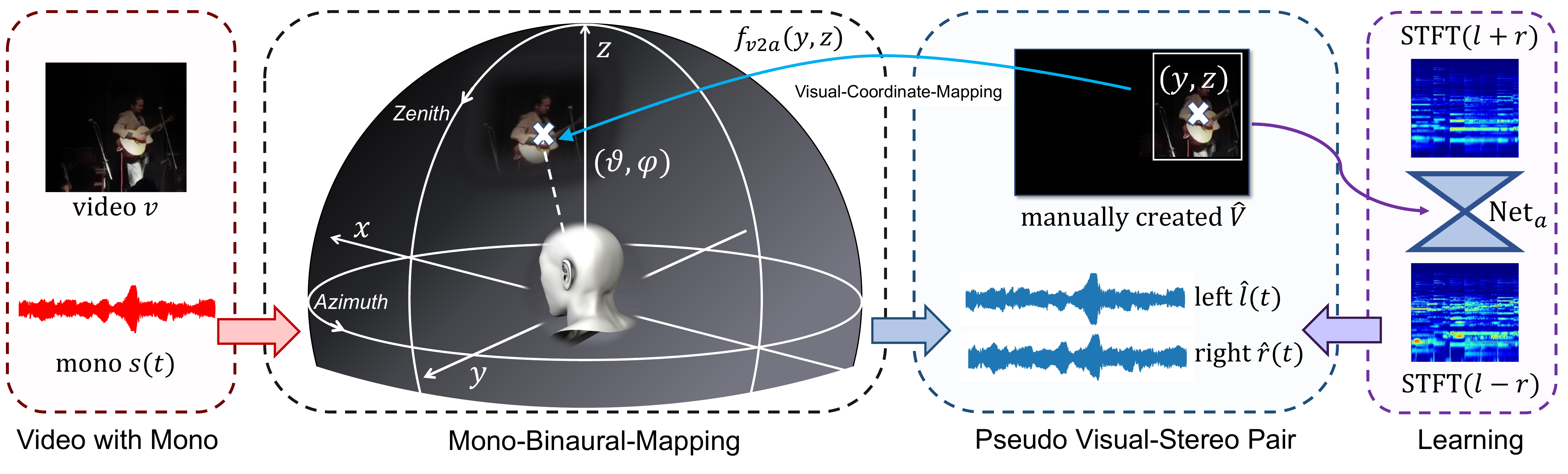}
    \vspace{-10pt}
    \caption{The pipeline of our method. Given one mono source, we create a \emph{pseudo visual-stereo} pair $\{\hat{V}, (\hat{l}, \hat{r})\}$ by assigning the source direction $\boldsymbol{\vartheta} = (\vartheta, \varphi)$ in the spherical coordinates according to our manually created $\hat{V}$. 
 Then mono source $s(t)$ is converted to binaural channels $(\hat{l}(t, \boldsymbol{\vartheta}), \hat{r}(t, \boldsymbol{\vartheta}))$ through our Mono-Binaural-Mapping procedure by leveraging spherical harmonics decomposition. 
Within this pipeline, multiple sources can be linearly blended together to build training pairs. Then mono-to-binaural networks can be trained on the created pseudo data. 
}
\vspace{-12pt}
\label{fig:teaser}
\end{figure*}

\section{Introduction}

Auditory and visual experiences are implicitly but strongly connected. 
In immersive environments, the perception of sound is impacted by visual scenes~\cite{zhou2019vision}. 
Therefore, researchers have explored ways to generate stereophonic audios
with visual guidance, in order to improve the user experience in multimedia products.
Specifically, supervised learning methods~\cite{morgado2018self,gao20192,lu2019self,zhou2020sep} 
have been considered for this purpose.


However, it is noteworthy that fully-supervised learning methods, despite the positive results that they achieve under constrained settings, would face significant difficulties
in real-world applications.
%
%
1) They rely on videos associated with stereophonic recordings, which we refer to as ``visual-stereo'' pairs~\cite{morgado2018self,gao20192}.
Obtaining a high-quality collection of real stereo data requires complicated and professional recording systems (\textit{e.g.} microphone arrays or dummy heads), thus is both resource-demanding and time-consuming.
2) The models trained on datasets collected under controlled environments may overfit to   
the layout of the rooms, rather than capturing the general associations between sound effects and
the visual locations of the sound sources.
The resultant models would also have poor generalization capability.

The privilege of learning representations from unlabeled data has been well discussed in different fields of deep learning~\cite{yang2020telling, korbar2018cooperative, alwassel2019self, morgado2020audio, owens2018audio}.
This inspires us to explore an alternative approach, namely, to use only mono audios which can be acquired much more easily compared to binaural audios.
%
%
We note that mono audios
have been successfully applied in learning visually informed sound separation~\cite{ephrat2018looking,afouras2018conversation,gao2018learning, zhao2018sound, zhao2019sound, gao2019co}. Zhou \textit{et al.}~\cite{zhou2020sep} recently leverage mono audios for stereo generation. 
However, their stereophonic learning procedure still depends on stereo data.

In this work, we propose \textbf{PseudoBinaural}, a novel pipeline that generates visually coherent binaural audios 
without accessing any recorded binaural data.
Our key insight is to carefully build \emph{pseudo visual-stereo} pairs from mono data.
Two questions need to be identified in order to achieve our goal.
Given a spatial location,
1) \emph{what is the relationship between a mono audio and its binaural counterpart sourcing from that location?} 2) \emph{How should visual cues be organized to represent the source visually?}
Our solution is to utilize two mappings. A \emph{Mono-Binaural-Mapping} to reproduce binaural audios of a single source positioned at any spatial location,
and a \emph{Visual-Coordinate-Mapping} that associates visual modality with spatial locations.
Specifically, 
the Mono-Binaural-Mapping is achieved by adopting spherical harmonic decomposition \cite{courant1962methods}. 
A head-related impulse response
(HRIR) \cite{begault20003} is then used to render binaural audios from the zero- and first-order terms of the decomposition.
As for the Visual-Coordinate-Mapping,
we pre-define a correspondence between pixel coordinates and spherical coordinates,
so that we can easily manipulate visual content to meet the designation of the corresponding source direction.

Existing models for visually informed binaural audio generation can be readily adapted to train on our \emph{pseudo visual-stereo} pairs. 
In order to make the best use of mono data, we further propose a new way of leveraging the task of audio-visual source separation~\cite{zhao2018sound,gao2019co} to assist the training.
The inference procedure is to simply apply the trained models to videos with mono audios and generate corresponding binaural audios.
%
%
%
Our framework renders stable performances on
two datasets
and in-the-wild scenarios.
Moreover, we can mix our \textit{pseudo} data with real stereophonic recordings to further boost the performance of binaural audio generation under the supervised setting.
%

%
%

Our contributions can be summarized as follows: 
\textbf{1)} We identify the mapping between source directions and binaural audios with theoretical analysis. 
\textbf{2)} By manipulating the visual modality, pseudo visual-stereo pairs can be generated for model training without relying on any recorded binaural data. 
\textbf{3)} Extensive experiments validate the effectiveness and stability of our method on a variety of scenes. Moreover, our pseudo visual-stereo data can serve as a strong augmentation under the supervised setting. 

\section{Related Work}
\label{sec:related}
\noindent\textbf{Visually Informed Stereophonic Audio Generation.}
%
While stereo is strongly correlated with visual information, only few papers have proposed to guide the generation of stereo with vision. 
Li \textit{et al.}~\cite{Li:2018:360audio} combine a synthesized early reverberation and a measured late reverberation tail for the generation of stereo sound in the desired room.
However, the usage of such method is restricted to specific rooms and serves for 360\degree videos. Morgado \textit{et al.}~\cite{morgado2018self} propose to recover ambisonics based on the datasets collected from YouTube.
They assume that their end-to-end network is able to separate sound sources and reformulate them with learnable weights. 
Lu \textit{et al.}~\cite{lu2019self} leverage flow with corresponding classifier for stereo generation. Specifically, Gao \textit{et al.}~\cite{gao20192} collect the FAIR-Play dataset using professional bianural audio collecting mics in a music room. Then they propose the Mono2bianural pipeline for converting mono audios to bianural ones in a U-Net framework. Their data is precious yet limited, models trained on their lab-collected data are difficult to generalize well on in-the-wild scenarios. Very recently, Zhou \textit{et al.}~\cite{zhou2020sep} leverage mono data and propose to tackle stereophonic audio generation and source separation at the same time. 
Nevertheless, their method uses mono data to train separation only.
All the above methods rely on recorded stereophonic data and visual-stereo pairs for training. We target to generate visually guided binaural audios without any binaural data.

\noindent\textbf{Visually Indicated Sound Source Separation.}
The task of visually guided sound source separation aims at separating a mixed audio into independent ones, according to their sound source's visual appearances.
It has long been an interest of research for both human speech~\cite{fisher2001learning,maganti2007speech,owens2016visually,ephrat2018looking,afouras2018conversation} and music~\cite{parekh2017motion,gao2018learning,zhao2018sound,zhao2019sound,gao2019co,gan2020music,tian2019deep,tian2021cyclic}. Recent learning-based methods~\cite{afouras2018conversation,zhao2018sound,zhao2019sound,gao2019co,xu2019recursive} all leverage the \textbf{Mix-and-Separate} training pipeline that creates training pairs using collected solo data. Our work also exploits the same type of data to build training samples for binaural generation. We also adopt the setting of separating two sources~\cite{zhou2020sep} to boost the final performance.

\noindent\textbf{Sound Source Localization.}
\label{sec:source}
One of the most important features for human auditory system is to localize sound by the subtle differences of intensity, spectral and time cues between ears~\cite{blauert1997spatial}. 
In the audio domain, previous research mostly relies on microphone arrays to perform direction of arrival estimation~\cite{zhang2012fast}.
Multi-modality works learn audio-visual associations~\cite{arandjelovic2017look,arandjelovic2018objects,tian2018audio,tian2020unified,zhou2019talking,wu2019dual,zhu2021learning} and propose to localize the responses of sound in the visual domains~\cite{zhao2018sound,senocak2018learning,qian2020multiple,hu2020discriminative}  a different type of ``source localization". 
Recent works~\cite{gan2019self,vasudevan2020semantic} propose to detect the position of vehicles with stereo audio, which deals with the opposite of our task. 
Normally, the visualization of activation is used to show auditorily associated visual information~\cite{arandjelovic2017look,owens2016visually,owens2018audio,arandjelovic2018objects,zhao2019sound,morgado2018self,gao20192}. 
Our model also shows the ability of source localization by training only on our pseudo visual-stereo pairs.

\begin{figure*}[t]
    \centering
    \includegraphics[width=0.97\linewidth]{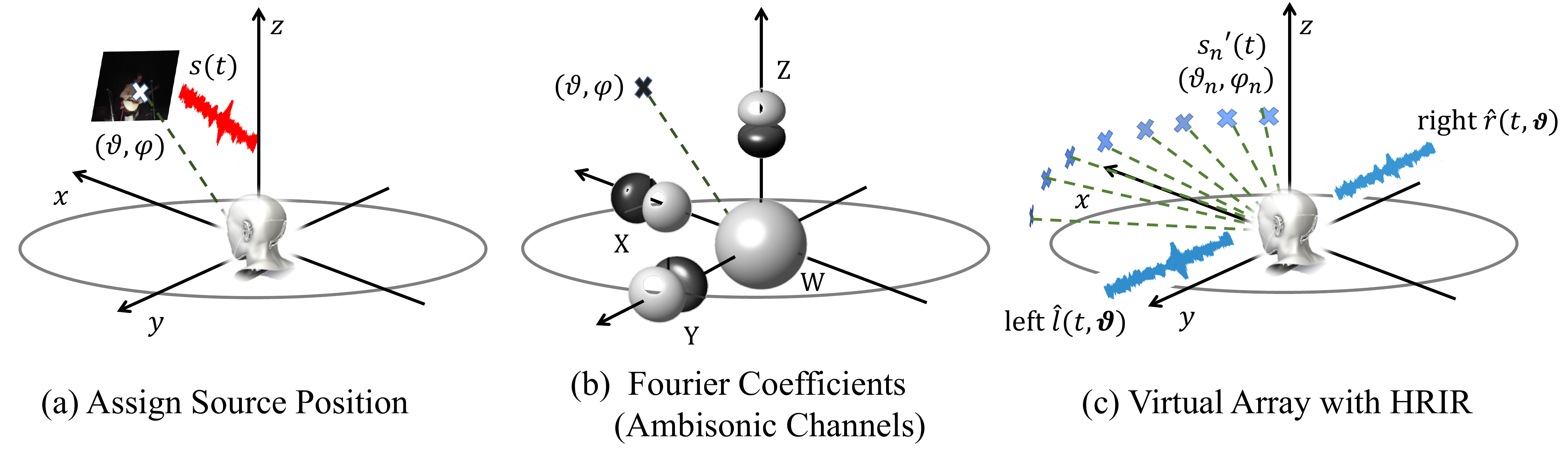}
    \vspace{-5pt}
    \caption{The steps for the Mono-Binaural-Mapping procedure. (a) Firstly the mono sound source $s(t)$ is assigned at direction $\boldsymbol{\vartheta} = (\vartheta, \varphi)$. (b) Then through spherical harmonics decomposition of the source, we can derive the zero- and first-order spherical-based Fourier coefficients (which are also ambisonic channels). The figure represents the directions of the channels. (c) Finally, the Fourier coefficients can be transferred to a set of speaker array with fixed positions, and generate binaural with HRIR.
}
\vspace{-10pt}
\label{fig:method}

\end{figure*}

\section{Methodology}

%
Different from previous completely learning-based and data-driven methods that rely on the ground-truth stereo, we train networks on self-created \textit{pseudo visual-stereo} pairs. Our method is thus called \textbf{PseudoBinaural}. The overall pipeline is illustrated in Fig. \ref{fig:teaser}. 

\subsection{Mapping Mono to Binaural}
\label{sec:3.1}
The key of our method relies on identifying the relationship between mono and stereo. This whole procedure, as illustrated in Fig.~\ref{fig:method}, is called \textit{Mono-Binaural-Mapping}. Given a mono audio with an arbitrarily assigned source position  $\boldsymbol{\vartheta} = (\vartheta, \varphi)$ (Fig.~\ref{fig:method}  \textbf{a}), our goal is to first convert it to binaural channels with correct auditory sense of location. 
We empirically choose spherical harmonic decomposition for its expressive ability and its substantial connection with ambisonics (Fig.~\ref{fig:method} \textbf{b}). Finally, the decomposed coefficients are transformed to virtual array and rendered to audios with HRIR (Fig.~\ref{fig:method} \textbf{c}). 

\noindent{\textbf{Spherical Harmonic Decomposition.}} 
The Laplace spherical harmonics represent a complete set of orthonormal basis defined on sphere surface~\cite{courant1962methods}. The normalized form of spherical harmonics defined at azimuth angle $\vartheta$ and zenith angle $\varphi$ in spherical coordinates can be represented as:
\begin{align}
\label{eq:1}
    Y^m_l(\varphi, \vartheta) = {N^{|m|}_l} P^{|m|}_l(\cos\varphi)e^{jm\vartheta},
\end{align}
where $P^{|m|}_l(\cos\varphi)$ is the associated Legendre polynomials, integer $l$ is its order and $m$ is the degree, limited to $[-l, l]$. ${N^{|m|}_l}$ is a normalization factor.
Real spherical harmonics can serve as a type of generalized Fourier series, to decompose any function $f$:
\begin{align}
\label{eq:2}
    f(\varphi, \vartheta) = \sum_{l=0}^{\infty} \sum_{m=-l}^{l} \Psi^m_l Y^m_l(\varphi, \vartheta).
\end{align}
The coefficients $\Psi^m_l$ are the analogs of Fourier coefficients which can be represented as ($^*$ denotes the conjunction):
\begin{align}
\label{eq:3}
    \Psi^m_l = \int_0^{2\pi} \int_0^\pi f(\varphi, \vartheta)Y^{m}_l(\varphi, \vartheta)^*\sin \varphi ~\mathrm{d}\varphi \mathrm{d}\vartheta.
\end{align}

\noindent{\textbf{Decomposed Coefficients for Mono Source.}} 
Here we follow the simplest assumption that only the impulse response from the direction $\boldsymbol{\vartheta} = (\vartheta, \varphi)$ of a single sound source $s(t)$ is received,  the Fourier coefficients can be derived from Eq.~(\ref{eq:3}) as: 
\begin{align}
\label{eq:4}
    \Psi^m_{l}(\boldsymbol{\vartheta}) = s(t)Y^{m}_l(\varphi, \vartheta).
\end{align}
This is the same as the encoding of ambisonics, where the $\Psi^m_{l}(\boldsymbol{\vartheta})$ can also be regarded as ambisonics' components. For simplicity, $\boldsymbol{\vartheta}$ is omitted in the following representations associated with this pre-defined direction.

The zero- and first-order components ($l=0, 1$) that contribute most to 3D audio effect are leveraged in our model. 
%
Based on Eq.~(\ref{eq:1}) and Eq.~(\ref{eq:4}), the coefficients $\{\Psi^0_0, \Psi^1_1, \Psi^{-1}_1, \Psi^0_1\}$ ($\boldsymbol{\vartheta}$ omitted) can be written as:
\begin{align}
\label{eq:5}
    &\Psi^0_{0}  = s(t){N^{0}_0}, \nonumber\\
    &\Psi^1_{1}  = s(t)  {N^{1}_1}\cos\varphi \cos\vartheta, \nonumber\\
    &\Psi^{-1}_{1}  = s(t)  {N^{1}_1}\cos\varphi \sin\vartheta, \nonumber\\
    &\Psi^{0}_{1} = s(t) {N^{0}_1}\sin\varphi,
\end{align}
which correspond to the W, X, Y and Z channels of ambisonics, respectively. W is the omnidirectional base channel, X, Y and Z are the orthogonal channels lie along 3D Cartesian axes as illustrated in Fig. \ref{fig:method} (b).

We adopt the Schmidt semi-normalization (SN3D)~\cite{winch2005geomagnetism} to Eq.~(\ref{eq:1}), which can be written as 
$N^m_l = \sqrt{(2 - \delta_m)\frac{(l-|m|)!}{(l + |m|)!}}$, where $\delta_m = 1$ if $m=0$ else $0$.

\noindent\textbf{Binaural Decoding.}
\label{sec:3.1.2}
Regarding the decomposed coefficients as ambisonic channels, we can roughly predict the left and right binaural channels $\hat{l}(t)$ and $\hat{r}(t)$ using simple transformation: $\hat{l}(t) = \text{W} + \text{Y}$ and $\hat{r}(t) = \text{W} - \text{Y}$. However, this paradigm is unable to recover real binaural.

On the other hand, binaural sound can be directly synthesized given a source position with the head-related impulse response (HRIR). 
One set of HRIR data can serve as filters $h_r({\boldsymbol{\vartheta}})$ and $h_l({\boldsymbol{\vartheta}})$ 
with respect to the direction $\boldsymbol{\vartheta}$ of the sound source. The transferred binaural sound can be represented as 
$\hat{l}(t) = h_l({\boldsymbol{\vartheta}}) \circledast  s(t)$ and $\hat{r}(t) = h_r({\boldsymbol{\vartheta}}) \circledast s(t)$,
where $\circledast $ is the convolution operation. 
However, open-sourced HRIR \cite{algazi2001cipic} are recorded in the free-field, thus cannot recover binaurals in a normal scene owing to reverberations.

Our solution is to leverage the binaural rendering technique that combines ambisonics with HRIR~\cite{noisternig20033d}. A virtual speaker array is pre-defined to make up for the reverberations as shown in Fig.~\ref{fig:method} (c). 
Denoting the Fourier coefficients (ambisonic channels)
as vectors $\boldsymbol{\Psi(\boldsymbol{\vartheta})} = s(t)\boldsymbol{Y}(\boldsymbol{\vartheta}) = [\Psi^0_{0}, \Psi^1_{1}, \Psi^{-1}_{1}, \Psi^0_{1} ]^{\text{T}}$ (Refer to Eq.~\ref{eq:5}), we can further decompose $\boldsymbol{\Psi(\boldsymbol{\vartheta})}$
 into $M$ virtual speakers at directions $\boldsymbol{\Theta}=[\boldsymbol{\vartheta}'_1, \dots, \boldsymbol{\vartheta}'_M]$ to analog the multi-source effect caused by room reverberations. We define the matrix $\boldsymbol{D(\boldsymbol{\Theta})} = [\boldsymbol{{Y}}(\boldsymbol{\vartheta}'_1), \dots, \boldsymbol{{Y}}(\boldsymbol{\vartheta}'_M)]$, each column representing the harmonics for each virtual source. The virtual audio signals $\boldsymbol{s'}(t) = [s'_1(t), \dots, s'_M(t)]^\text{T}$ can be constraint as: 
\begin{align}
\label{eq:6}
    \boldsymbol{D(\boldsymbol{\Theta})}\boldsymbol{s'}(t) = \boldsymbol{\Psi(\boldsymbol{\vartheta})}, 
\end{align}
As the matrix $\boldsymbol{D(\boldsymbol{\Theta})}^\text{T}\boldsymbol{D(\boldsymbol{\Theta})}$ is of full-column rank, the virtual signals can be computed:
\begin{align}
\label{eq:7}
    \boldsymbol{s'}(t) =(\boldsymbol{D(\boldsymbol{\Theta})}^\text{T}\boldsymbol{D(\boldsymbol{\Theta})})^{-1}\boldsymbol{D(\boldsymbol{\Theta})}^\text{T} \boldsymbol{\Psi(\boldsymbol{\vartheta})}.
\end{align}
Finally, we take advantage of HRIR filters to acquire the desired left and right channels:
\begin{align}
\label{eq:8}
    &\hat{l}(t, \boldsymbol{\vartheta}) = \sum_{m=1}^{M} h_l({\boldsymbol{\vartheta}'_m}) \circledast  s'_m(t), \nonumber\\
    &\hat{r}(t, \boldsymbol{\vartheta}) =  \sum_{m=1}^{M} h_r({\boldsymbol{\vartheta}'_m}) \circledast  s'_m(t).
\end{align}
The above is the binaural audio generated from a single mono source $s(t)$ from direction $\boldsymbol{\vartheta}$.

\begin{figure*}[h]

    \centering
    \vspace{-10pt}
    \includegraphics[width=0.97\linewidth]{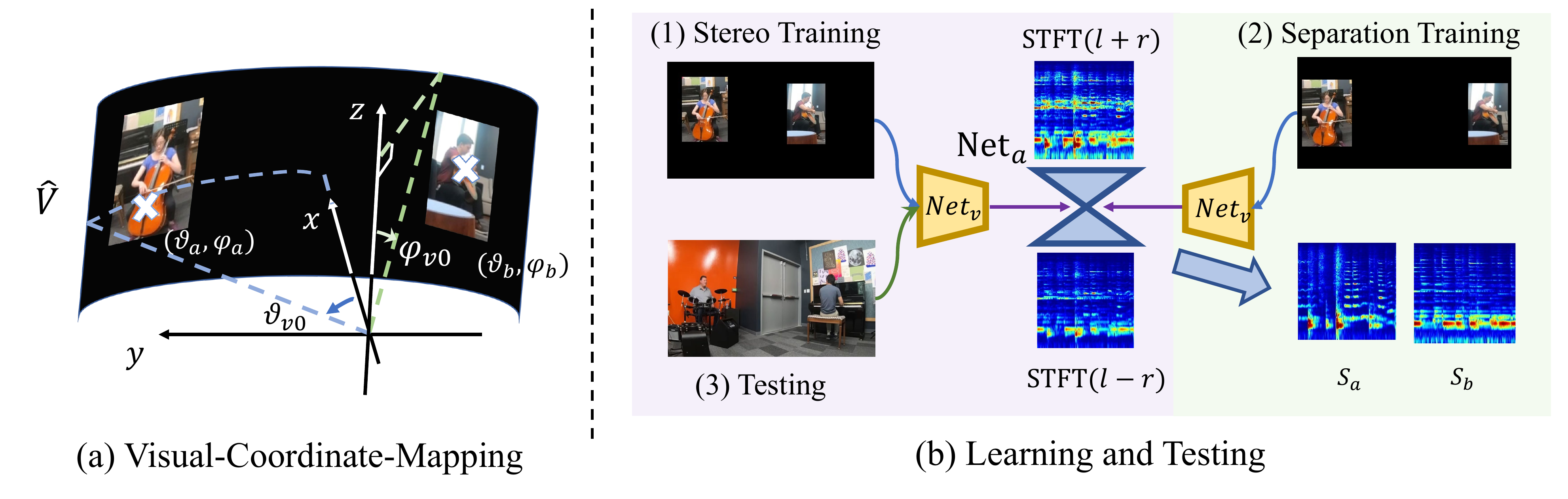}
    \caption{(a) Mapping from visual positions in the image domain to spherical angles. Normally we place the image $\hat{V}$ at the frontal-view to be part of a cylinder. The borders of the image are corresponding to the angles $\varphi_{v0}$ and $\vartheta_{v0}$. (b) Network details. The input to the audio UNet $\text{Net}_a$ is the STFT of a mono audio. The output is the prediction of the STFT difference between the left and right channels. During training, $\text{Net}_v$ extract the visual feature $f_v$ from our self-created image $\hat{V}$, and concatenate it to the audio UNet. During testing, this network can be applied to normal frames.
}
\vspace{-5pt}
\label{fig:learning}
\end{figure*}

\subsection{Creating Pseudo Visual-Stereo Pairs}
\label{sec:3.2}
With the Mono-Binaural-Mapping. there are two questions left to achieve visually informed binaural generation: 1) How to leverage the visual information, and 2) how to connect the direction of the sound source with visual information. To this end, we create \textit{pseudo} visual information and define a \textit{Visual-Coordinate-Mapping}. 

\noindent\textbf{Pseudo Visual Information Creation.} As illustrated in Fig.~\ref{fig:teaser}, assuming that the listener is facing towards the $x$ axis in 3D Cartesian coordinates, frontal-view scenes can thus be projected to the $y-z$ image plane. 
Given a video $v_k$ with a single sound source, we can place the center of the sound source $v'_k$ to a random position in the image plane. More specifically, the cropped frames $v'_k$ are placed to a background image $\hat{V}$ according to $\hat{V}(y, z) = v'_k$.  

\noindent\textbf{Visual-Coordinate-Mapping.} 
\label{sec:visual-}
%
We then define the mapping $f_{v2a}$ from pixel position $(y, z)$ on frontal-view images to spherical angles $\boldsymbol{\vartheta} = f_{v2a}(y, z)$. 
In the spherical coordinate, the frontal-view image plane is defined as part of a cylinder centered at the coordinate origin as shown in Fig.~\ref{fig:learning} (a). 

Based on the fact that the effective visual field of humans is approximately 120 degrees~\cite{smythies1996note},
we define the border azimuth angle as $\vartheta_{v0} = \pi/3 $. So that objects within $\hat{V}$ are distributed within  $ \vartheta \in [-\vartheta_{v0}, \vartheta_{v0}]$. 
The ratio between the height and width of the background image is set to 
$H/W = 1/2$, thus the top edge of the image is corresponding to $\varphi_{v0} = \pi/2 - \arctan(\pi / 3) $. The range is $\varphi \in [\pi/2 - \arctan(\pi / 3), \pi / 2 + \arctan(\pi / 3)]$. In this way, for each point in the image plane, we can find an angle in the spherical coordinate. We have also explored other field of view settings and find subtle differences.

\noindent\textbf{Create Pseudo Visual-Stereo Pairs.}
By calculating the corresponding angle $\boldsymbol{\vartheta}_k$ in spherical coordinates,
a pair of binaural audios $\{\hat{r}_k(t, \vartheta_k), \hat{l}_k(t, \vartheta_k)\}$ can be recovered from the mono audio $s_k(t)$ accompanying $v_k(t)$  through Eq.~(\ref{eq:8}).

Audio recordings collected in real-world scenarios are mostly mixed,
therefore, we propose to mix multiple solo videos together in one scene to create \textit{pseudo visual-stereo} pairs  $\{\hat{V}, (\hat{l}, \hat{r})\}$. 
Each time, we assemble a random number of $K$ independent mono videos $\boldsymbol{s}(t) = \{s_1(t), \dots, s_K(t)\}$ together to form a pseudo visual-stereo pair. The self-created binaural can be written as $\hat{l}(t) = \sum_{k=1}^{K}\hat{l_k}(t, \vartheta_k)$ and $\hat{r}(t) = \sum_{k=1}^{K}\hat{r_k}(t, \vartheta_k)$. The manually built visual information is $\hat{V}$, where $\hat{V}(\boldsymbol{\vartheta}_k) = v_k, k \in [1, K]$.

Note that the patch size and audio amplitude are both directly proportional to the reciprocal of the depth. Therefore, the mono audio is firstly normalized according to its wave amplitude, and the corresponding cropped patch $v_k'$ is normalized in the same scale as the audio. When the pseudo scene is assembled, $v_k'$ is randomly resized and placed on $\hat{V}$ to represent objects at different depths. Only mono audios are used in building the data. 

\subsection{Learning}
\label{sec:3.3}
We leverage neural networks for learning from mono and visual guidance to pseudo binaural output. Previous networks and training paradigms from Mono2Binaural~\cite{gao20192} and Sep-Stereo~\cite{zhou2020sep} can be readily adapted to train on our data. The learning procedure is depicted in Fig.~\ref{fig:learning} (b).

\noindent\textbf{Stereo Training.} \label{sec:3.3.2}
The main part of our learning procedure is to directly train networks using our pseudo visual-stereo pairs.
The whole training procedure is basically following~\cite{gao20192,zhou2020sep}. It consists of a backbone U-Net~\cite{ronneberger2015u} audio network $\text{Net}_a$, and a ResNet18~\cite{he2016deep} visual network $\text{Net}_v$. 
The audios are all processed in the complex Time-Frequency domain in form of Short-Time Fourier Transformation (STFT). 
Mono is created from the left and right channels $s_m(t) = \hat{l}(t) + \hat{r}(t)$ and the input to $\text{Net}_a$ is the transformed mono spectrum $S_m = \text{STFT}(s_m(t))$. 
$\text{Net}_a$ returns the complex mask $\mathcal{M}$ for final predictions. The training objective is the difference of the left and right spectrums $S_D = \text{STFT}(\hat{l}(t) - \hat{r}(t))$, which can be written as:
\begin{align}
\label{eq:9}
    \mathcal{L}_{stereo} = \| S_D - \mathcal{M} * S_m\|_2
\end{align}
Then the predicted difference spectrum is transferred back to the difference audio ${\widetilde{s}_D} = \text{ISTFT}(\mathcal{M} * S_m)$ . The predicted left and right can be computed as $\widetilde{l}(t) = (s_m(t) + \widetilde{s}_D(t))/2$ and $\widetilde{r}(t) = (s_m(t) - \widetilde{s}_D(t))/2$.

\noindent\textbf{Separation Training.} Specifically, we leverage the task of separating two sources inspired by Sep-Stereo~\cite{zhou2020sep}. 
We care less about the performance on separation, but more about its benefits on distinguishing sound sources.
Thus different from their visual feature rearrangement, we directly place two sources at separate edges when creating the pseudo visual input $\hat{V}$ (as shown in Fig.~\ref{fig:learning} (b)). The network input would be the pseudo pair $\{\hat{V}, (s_a, s_b)\}$, where $s_a$ and $s_b$ are the individual mono audio signals.
%
Then we leverage one APNet branch from~\cite{zhou2020sep} to predict the original STFTs $S_a$ and $S_b$. In this way, the backbone network can learn better the association between the sources' visual and audio information. Please refer to the supplementary materials for details.
\begin{table*}[t] 
\begin{center}
\caption{Quantitative results of binaural audio generation on FAIR-Play and MUSIC-Stereo dataset.
Except for SNR, the lower the score, the better the results. The upper half shows the results of standard benchmarks and our PseudoBinaural method. 
%
%
The lower half shows the augmentation results and our ablation studies on different binaural decoding schemes (Sec.~\ref{sec:3.1.2}). 
Our method outperforms previous methods when augmented with binaural recordings.
Moreover, our chosen decoding scheme  achieves the best performance among three decoding methods. 
}
\label{table:exp1}
\begin{tabular}{lcccccccccc}
\toprule
 & \multicolumn{5}{c}{FAIR-Play }& \multicolumn{5}{c}{MUSIC-Stereo} \\
\cmidrule(lr){2-6} \cmidrule(lr){7-11}
Method & $\text{STFT} $ & $\text{ENV} $ & $\text{Mag} $ & $\mathcal{D}_{phase} $ & $\text{SNR} \uparrow $ & $\text{STFT} $ & $\text{ENV}$ & $\text{Mag} $ & $\mathcal{D}_{phase} $  & $\text{SNR} \uparrow$ \\

\midrule
Mono-Mono &1.024 &0.145 &2.049 &1.571 &4.968 &1.014 &0.144 &2.027 &1.568 &7.858 \\
Mono2Binaural~\cite{gao20192}  &0.917 &0.137 &1.835 &1.504 &5.203 &0.942 &0.138 &1.885 &1.550 &8.255 \\

\textbf{PseudoBinaural} (w/o sep.) &0.951 &0.140 &1.914 &1.539 & 5.037 &0.953 &0.139 &1.902 &1.564 &8.129 \\
\textbf{PseudoBinaural} (Ours) & 0.944 & 0.139 & 1.901 & 1.522 & 5.124 &0.943 &0.139 &1.886 &1.562 &8.198 \\
\hline
\hline
Sep-Stereo~\cite{zhou2020sep} &0.906 &0.136 &1.811 &1.495 &5.221 &0.929 &0.135 &1.803 &1.544 &8.306 \\
Augment-HRIR  &0.896 &0.137 &1.791 &1.472 &5.255 &0.940 &0.138 &1.866 &1.550 &8.259 \\
Augment-ambisonic &0.912 &0.139 &1.823 &1.477 &5.220 &0.909 &0.137 &1.817 &1.546 &8.277 \\
\textbf{Augment-PseudoBinaural} &\textbf{0.878} &\textbf{0.134} &\textbf{1.768} &\textbf{1.467} &\textbf{5.316} &\textbf{0.891} &\textbf{0.132} &\textbf{1.762} &\textbf{1.539} &\textbf{8.419} \\

\bottomrule
\end{tabular}
\end{center}

\vspace{-10pt}
\end{table*}

\section{Experiments}


\subsection{Datasets} 

We emphasize creating binaurals for music, which is an important scenario for stereo production. We will at first show our analysis on the FAIR-Play dataset, then introduce other datasets we use.

\noindent\textbf{Revisiting FAIR-Play.}
Collected in a music room, FAIR-Play~\cite{gao20192} is one of the most influential datasets in this field. 
However, by carefully examining the dataset, we find that the original train-test splits are somewhat problematic. The whole dataset contains $1,871$  clips cut from several different long camera recordings with  approximately the same camera view and scene layouts. 
The clips are randomly divided into 10 different train-test splits. As a result, the scenes within train and test splits are overlapped, probably originate from the same recording. 
This would lead to serious overfitting problems. The models might learn layouts of the room instead of visual-stereo association that we desired.


In order to evaluate the true generalization ability of different models on this dataset, we take efforts to re-organize the FAIR-Play dataset through reconstructing the original videos and re-splitting them.
%
Specifically, we first run a clustering algorithm on all the clips to roughly group them according to the scenes. Then by matching the first and last frame of each clip within groups, we find the original order of the clips and concatenate them to recover the recorded videos. Finally, we select the videos whose scenes are completely absent in other videos as the validate and test sets.
In this way, we create 5 different splits in which train and test sets are not overlapped.
In our experiments, we re-train all supervised models (including Mono2Binaural~\cite{gao20192} and Sep-Stereo~\cite{zhou2020sep}) and report the average results on the five splits. Please be noted that our model is also trained on this dataset, using only the solo part and \emph{mono audios}.

\noindent\textbf{MUSIC-Stereo}~\cite{zhao2018sound,zhao2019sound}. 
Containing 21 different types of instruments, MUSIC(21) is originally collected for visually guided sound separation. We select all the videos with binaural audio from MUSIC(21) and MUSIC-duet~\cite{zhao2018sound}
to form a new dataset MUSIC-Stereo. Composed of solo and duet parts, it includes $1,120$ unique videos of different musical performances. MUSIC-Stereo lasts 49.7 hours in total, which is 10 times larger than the FAIR-Play dataset. 
Following the post-processing steps in \cite{gao20192}, we cut these videos into $17,940$ 10s clips and split them into training, validation, and test sets in an 8:1:1 ratio. Similar to FAIR-Play, only the solo part and \emph{mono audios} are exploited for our model's training. 

\noindent\textbf{YT-Music}~\cite{morgado2018self}. This dataset is collected from 360\degree videos on YouTube in the ambisonic format. 
The audios are transferred to binaural in the same way as our decoding scheme. 
With distinct vision configurations and stereo audio characteristics, YT-MUSIC is the most challenging dataset.



\begin{table*}[t] 
\setlength{\tabcolsep}{8pt}
\vspace{-10pt}
\begin{center}
\caption{Cross-dataset evaluation results on five metrics. While the model trained on FAIR-Play is used for testing on the others, the model trained on MUSIC-Stereo is for the evaluation on FAIR-Play. PseudoBinaural presents better generalization ability than the supervised method Mono2Binaural on all datasets.
}
\label{table:exp2}
\begin{tabular}{lcccccccccc}
\toprule
 & \multicolumn{5}{c}{Mono2Binaural~\cite{gao20192}} & \multicolumn{5}{c}{\textbf{PseudoBinaural (Ours)}} \\
\cmidrule(lr){2-6} \cmidrule(lr){7-11}
Dataset & $\text{STFT} $ & $\text{ENV} $ & $\text{Mag} $ & $\mathcal{D}_{phase} $ & $\text{SNR} \uparrow $ & $\text{STFT} $ & $\text{ENV}$ & $\text{Mag} $ & $\mathcal{D}_{phase} $  & $\text{SNR} \uparrow$ \\

\midrule
FAIR-Play  &0.996 &0.142 &1.993 &1.562 &5.876 &\textbf{0.959} &\textbf{0.140} &\textbf{1.917} &\textbf{1.496} &\textbf{6.057} \\
MUSIC-Stereo  &0.971 &0.140 &1.942 &1.552 &7.933 &\textbf{0.952} &\textbf{0.139} &\textbf{1.904} &\textbf{0.574} & \textbf{8.099}\\
YT-MUSIC  &0.717 &0.118 &1.435 &1.597 &9.214  &\textbf{0.653} &\textbf{0.111} &\textbf{1.306} &\textbf{1.357} & \textbf{9.848} \\

\bottomrule
\end{tabular}
\end{center}
\vspace{-15pt}
\end{table*}

\subsection{Evaluation Metrics}
\noindent\textbf{Previous Metrics.} 
The evaluation protocol within this field is basically the STFT distance and the envelope distance (ENV) between recovered audios and recorded ones~\cite{morgado2018self,gao20192}. The STFT distance represents the mean square error computed on predicted spectrums, and the ENV distance is performed on raw audio waves through Hilbert transform~\cite{smith2007mathematics}. To evaluate the predicted binaural audios more comprehensively, we also adopt two widely-used metrics \emph{Magnitude Distance (Mag)} and \emph{Signal-to-noise Ratio (SNR)} from \cite{morgado2018self}. The Mag distance reflects the $L2$ deviation on the magnitude of spectrums and SNR is operated on the waveform directly. 

\noindent\textbf{Newly Proposed Metric.} 
In 3D audio sensation, audiences care more about sensing the source direction, where the phase of binaural audio is the key.
As illustrated in \cite{neurals}, phase errors will introduce perceivable distortions but are always neglected during the optimization. Inspired by this, we further propose a new metric named \textit{Difference Phase Distance} ($\mathcal{D}_{phase}$), which is performed on the Time-Frequency domain. Note that, the binaural audio is completely determined by the difference between left and right spectrums~\ref{sec:3.3.2}. Hence, $\mathcal{D}_{phase}$ is to evaluate the phase distortion between the ground-truth difference $S_D$ and the predicted one $\Tilde{S_D} = \mathcal{M} * S_m$:
\begin{align}
\label{eq:10}
    \mathcal{D}_{phase} = \| \text{phase}(S_D) - \text{phase}(\Tilde{S_D}) \|_2,
\end{align}
where the phase is represented by the angle values, thus $\mathcal{D}_{phase} \in [0, 2\pi]$. It's worth emphasizing that $\mathcal{D}_{phase}$ is sensitive to the audio directions, \ie, switching left and right channels would bring a significant change on this metric.

\subsection{Quantitative Results}
\noindent\textbf{Binaural-Recording-Free Evaluation.} 
Since no binaural-recording-free method has been proposed before, supervised method \emph{Mono2Binaural}~\cite{gao20192} whose backbone we borrow from, can be served as our baseline and upper bound. The evaluation is made on our newly-split FAIR-Play~\cite{gao20192} and MUSIC-Stereo.
For comparison, \emph{Mono2Binaural} is trained with both visual frames and binaural audio, whereas our method \emph{PseudoBinaural} only leverages frames and mono audio to do the training. Please be noted that we do not rely on an extra dataset.
The result of \emph{Mono-Mono} is also listed, which copies the mono input two times as the stereo channels. This method should have no stereo effect at all, thus outperforming it means the success of generating the sense of directions.
As the whole model of ours includes the separation training described in Sec.~\ref{sec:3.3.2}, we also evaluation the ablation of this module \textbf{(w/o sep.)}.
Table~\ref{table:exp1} shows the results of these methods on all metrics.


With no supervision, it is reasonable that our \emph{PseudoBinaural} cannot outperform the supervised setting. However, the fact that our model outperforms \emph{Mono-Mono} on all metrics proves the effectiveness of our proposed method. In line with previous work~\cite{zhou2020sep}, introducing the separation task in the training framework can further improve the overall performance of generated binaural audio.

\noindent\textbf{Augmentation to Binaural Audio Training.} 
Since our method just relies on the pseudo visual-stereo pairs, a natural idea is to leverage both pseudo data and recorded ones to boost the performance of the traditional fully-supervised approach. As demonstrated in the lower half of Table~\ref{table:exp1}, our method, denoted as \emph{Augment-PseudoBinaural}, can surpass the traditional setting \emph{Mono2Binaural}~\cite{gao20192} on all 5 metrics. 
Moreover, compared to \emph{Sep-Stereo}~\cite{zhou2020sep}, which incorporates extra data, we create pseudo pairs with the same set of collected data,
providing more effective and complementary information to guide the training.
Consequently, our method outperforms theirs on both FAIR-Play and MUSIC-Stereo.
%


\noindent\textbf{Cross-Dataset Evaluation.} 
We specifically show the results of cross-dataset evaluation in Table.~\ref{table:exp2} to prove that 1) supervised methods can easily overfit to a specific domain and 2) the generalization ability of our method. YT-MUSIC~\cite{morgado2018self} with special 360\degree videos and ambisonics sounds is also used for evaluation. Here we use the non-augmented version of \emph{PseudoBinaural} for evaluation.
For \emph{Mono2Binaural}, the model evaluated on FAIR-Play is trained on MUSIC-Stereo, and the model tested on MUSIC-Stereo and YT-MUSIC is trained on FAIR-Play.
During cross-testing on FAIR-Play and MUSIC-Stereo, the visual to angle mapping $f_{v2a}$ is defined in the frontal view. But when cross-testing on YT-MUSIC, the video is defined in the form of 360\degree. 

We can see from the table that our method stably outperforms \emph{Mono2Binaural} in all cross-dataset evaluations. The supervised method tends to perform badly when testing in a different domain, while our recording-free method generalizes well by training only on mono data.


\begin{table}[t] 
\setlength{\tabcolsep}{8pt}
\begin{center}
\caption{Ablation study on the number $K$ of mono videos to mix based on FAIR-Play dataset. 
When $K$ is a mixture of the three different numbers, the ratio is empirically set to $1:2:3 = 0.4:0.5:0.1$.
}
\label{table:exp3}
\begin{tabular}{lccccc}
\toprule
$K$ & $\text{STFT} $ & $\text{ENV} $ & $\text{Mag} $ & $\mathcal{D}_{phase} $ & $\text{SNR} \uparrow $ \\
\midrule
1  &0.965 &0.143 &1.914 &1.483 &4.976 \\
2  &0.935 &0.141 &1.871 &1.480 &5.026 \\
3  &0.967 &0.142 &1.936 &1.527 &5.004 \\
1,2 &0.895 &0.136 &1.793 &1.479 &5.282 \\
1,2,3  &\textbf{0.878} &\textbf{0.134} &\textbf{1.768} &\textbf{1.467} & \textbf{5.316} \\

\bottomrule
\end{tabular}
\end{center}
\vspace{-18pt}
\end{table}


\begin{figure*}[!t]
    \centering
        \subfloat[User Study]{
        \begin{minipage}[b]{.45\textwidth}
        \flushright
        \includegraphics[width=1\textwidth]{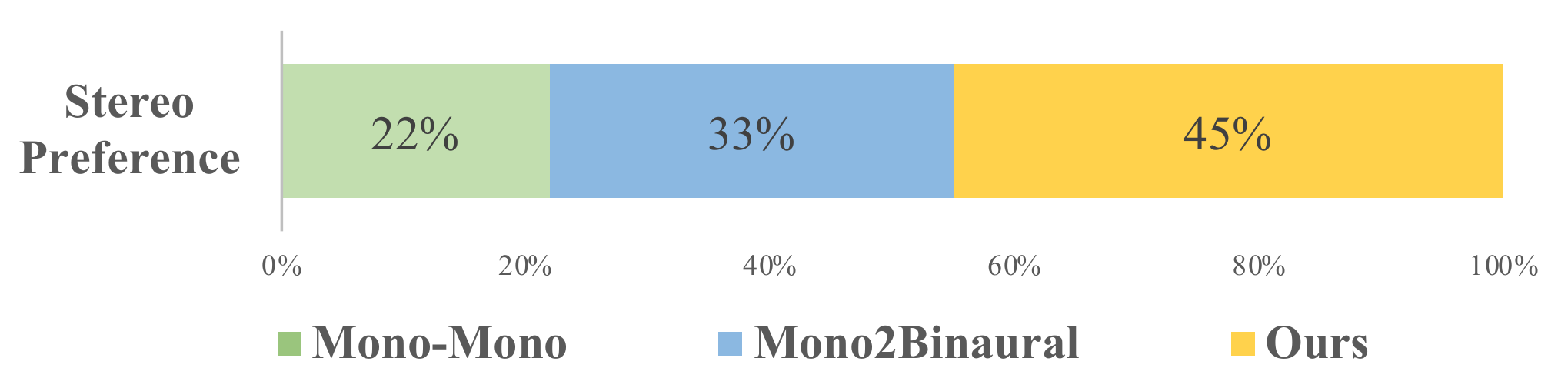}\\
        \includegraphics[width=1\textwidth]{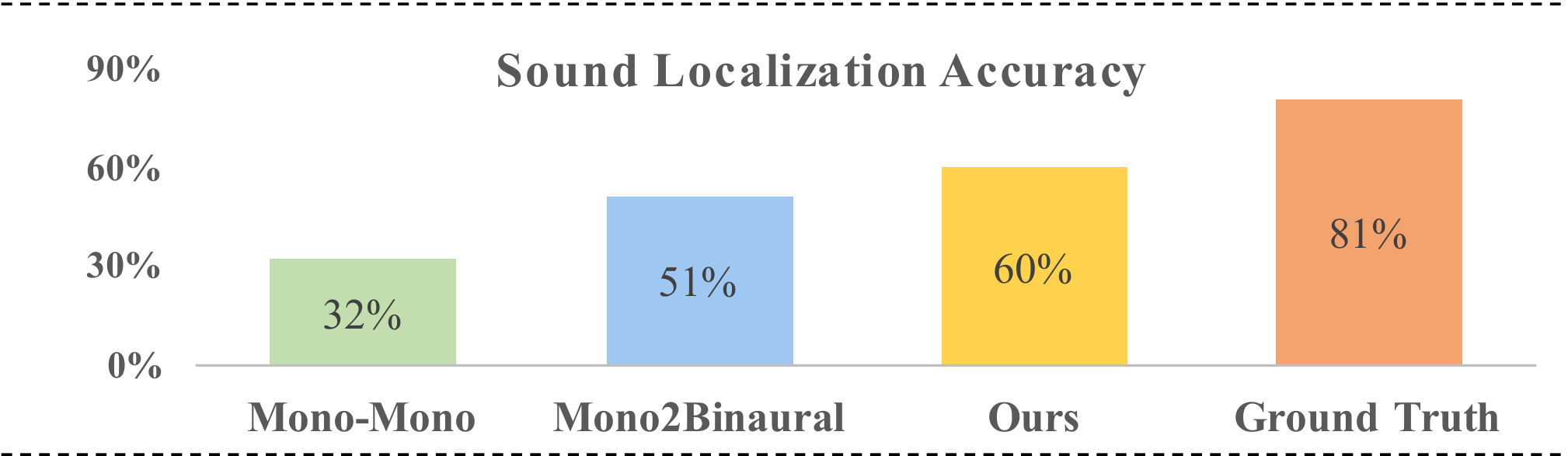}\\
        \includegraphics[width=1\textwidth]{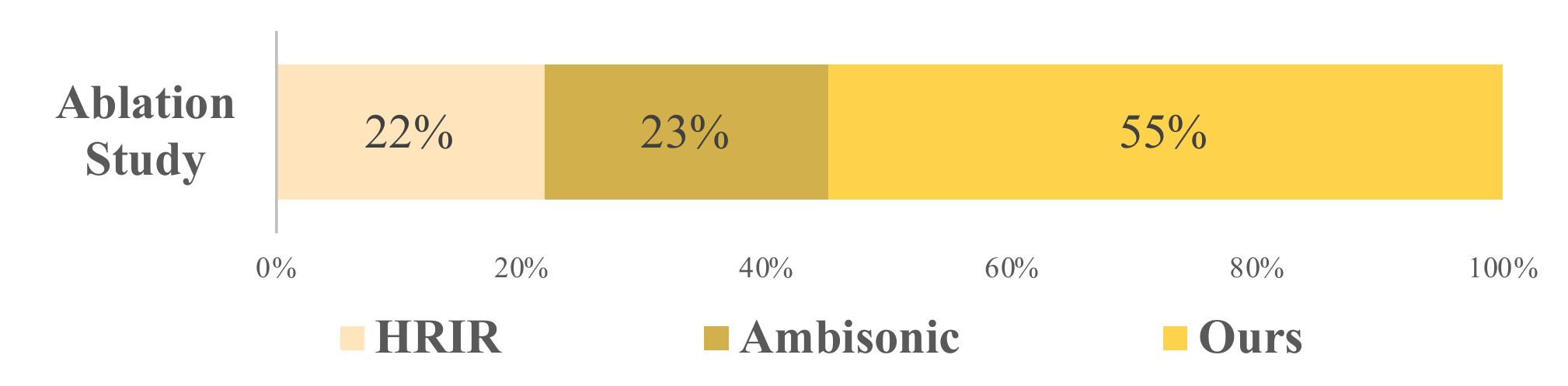}
        \end{minipage}
        }
        \subfloat[Activation Map]{
        \begin{minipage}[b]{.25\textwidth}
        \includegraphics[width=1\textwidth]{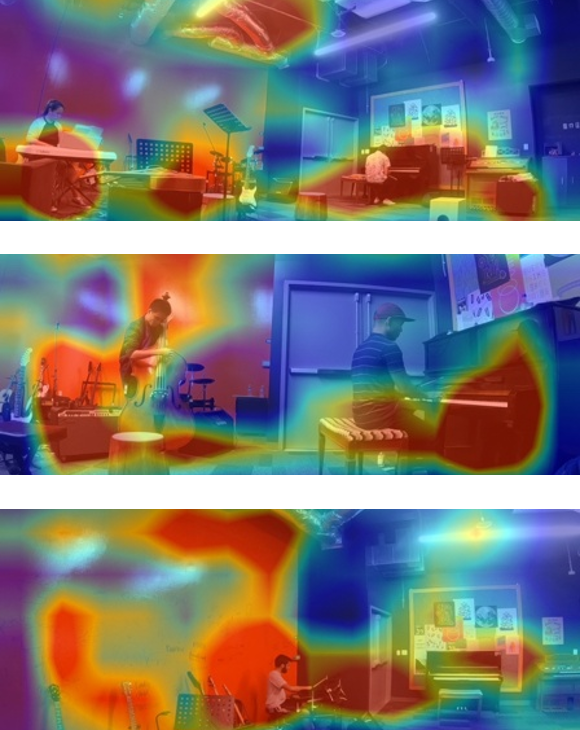}
        \centerline{\small{Mono2Binaural~\cite{gao20192}}}
        \end{minipage}
        \begin{minipage}[b]{.25\textwidth}
        \includegraphics[width=1\textwidth]{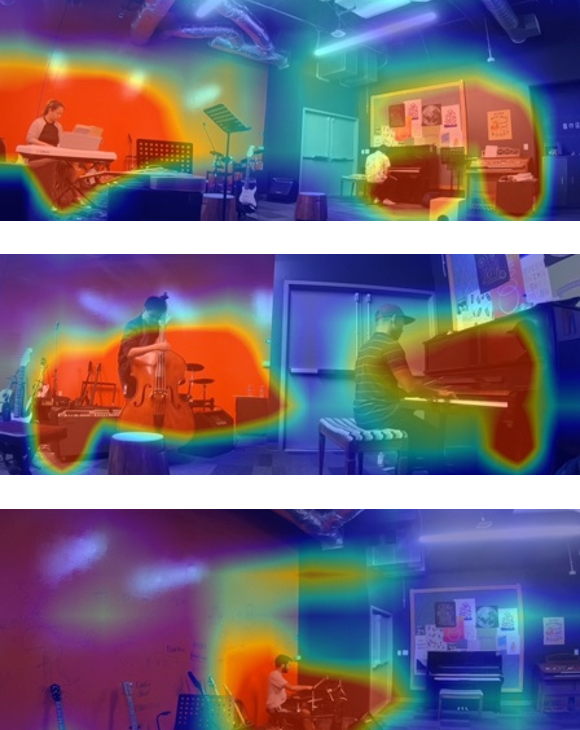}
        \centerline{Ours}
        \end{minipage}
        }
	\caption{
		Qualitative results. (a) shows the results of our user studies. It can be seen that the users slightly prefer our approach over supervised Mono2Binaural~\cite{gao20192}. (b) is the visualization of the activation maps of ours and Mono2Binaural. While the attention of theirs is messier, the results of ours are more compact. We focus more on sound sources. 
	}
    \label{fig:localization}
    \vspace{-5pt}
\end{figure*}

\noindent\textbf{Ablation Study.} 
The lower half of Table~\ref{table:exp1} presents our ablation studies on different binaural decoding schemes. 
As written in Sec.~\ref{sec:3.1.2}, binaural audios can be decoded directly from \emph{HRIR} or \emph{ambisonic}. It can be seen that our way of combining both leads to the best results.

When preparing pseudo visual-stereo pairs, the number of mono videos to mix is also another important hyperparameter for consideration. As shown in Table~\ref{table:exp3}, a fixed mixing number $K$ always fails to construct various training samples, introducing inconsistency with those naturally collected datasets. 
Hence, an empirical ratio of $1:2:3 = 0.4:0.5:0.1$ for the number $K$ is applied to ensure the diversity of generated visual-stereo pairs.

Additionally, we evaluate the choice of visual field-of-view (FOV)  when building the \emph{Visual-Coordinate-Mapping}. The influential parameter is the border azimuth angle $\vartheta_{v0}$ which is set to $\pi/3$ (Sec.~\ref{sec:visual-}). The results are shown in Table~\ref{table:FOV}, that our choice achieves the best results.

\subsection{Qualitative Results}
\noindent\textbf{User Study\footnote{Please refer to \url{https://sheldontsui.github.io/projects/PseudoBinaural} for demo videos.}.} 
In total 30 users with normal hearing participated in our study
to perform the quality evaluation.  There are three sets of studies, each with 20 videos selected from the test set of FAIR-Play~\cite{gao20192} and MUSIC-Stereo~\cite{zhao2019sound}, most of which are duets. 1) The users are asked to watch one video and listen to the binaural audios generated by \emph{PseudoBinaural}, \emph{Mono2Binaural}~\cite{gao20192} or \emph{Mono-Mono}. The question is which one of the three methods creates the best stereo sensation. The results show the percentage of the users' \textbf{Stereo Preferences.} 2) The users are asked to listen to the audio generated by the above methods without viewing videos, and decide where is the specific instrument (left, right, or center). Ground truth audios are also included for reference. The results show the \textbf{Sound Localization Accuracy} of these methods. 3) A subjective \textbf{Ablation Study} is conducted to show the influence of different choices of binaural decoding methods. The users are asked to tell which decoding diagram, direct HRIR, ambisonic, or ours, creates the best 3D hearing experience. 

The results are shown in Fig~\ref{fig:localization} (a). From the first and second experiments, we can see that users find our method slightly better than supervised \emph{Mono2Binaural} on both the two measurements.
This is enough to validate that our results are highly competitive to supervised methods in subjective evaluations, which is extremely important for auditory tasks. 
In the sound localization experiment, users can only achieve $81\%$ accuracy even given the ground-truth audio, which demonstrates the misguiding caused by the room reverberations.
The subjective ablation study shows that our decoding procedure apparently creates the best sense of hearing among all decoding choices.

\begin{table}[t] \footnotesize

\setlength{\tabcolsep}{9pt}
\caption{Ablation study on the border azimuth angle $\vartheta_{v0}$. The horizontal visual field-of-view is $2\vartheta_{v0}$.}
\vspace{-20pt}

\label{table:FOV}
\begin{center}
\begin{tabular}{lccccc}
\toprule
$\vartheta_{v0}$ & $ \pi/6 $ & $\pi/4 $ & $\pi/3 $ & $5\pi/12 $ & $ \pi/2$ \\
\midrule
STFT $\downarrow$  &0.923 &0.896 &\textbf{0.878} &0.884 &0.886 \\
SNR $\uparrow$  &5.138 &5.181 &\textbf{5.316} &5.302 &5.271 \\

\bottomrule
\end{tabular}
\end{center}
\vspace{-20pt}
\end{table}

\noindent\textbf{Visualization.} 
We also visualize the activation map generated by our method and \emph{Mono2Binaural}~\cite{gao20192} on the visual domain.
In Fig.~\ref{fig:localization} (b) we can see that \emph{PseudoBinaural} can successfully attend to sound sources while \emph{Mono2Binaural}~\cite{gao20192} would focus on less important areas. For example, their approach would attend to the ceiling for all three scenes shown, which is not the sound source. 




\section{Conclusion}

In this work, we propose PseudoBinaural, a binaural-recording-free method for generating binaural audios from corresponding mono audios and visual cues. 
%
For the first time, the problem of visually informed binaural audio generation is tackled without binaural audio recordings.
Based on the theoretical analysis of Mono-Binaural-Mapping, the created pseudo visual-stereo pairs can be capitalized to train models for binaural audio generation.
Extensive experiments validate that our framework can be very competitive both quantitatively and qualitatively. More impressively, augmented with real binaural audio recordings, our PseudoBinaural could outperform current state-of-the-art methods on various standard benchmarks. \\

\vspace{-5pt}
\noindent\textbf{Acknowledgements.} 
This work is supported by the Collaborative Research Grant from SenseTime (CUHK Agreement No. TS1712093), the General Research Fund (GRF) of Hong Kong (No. 14205719,  14202217, 14203118, 14208619), NTU NAP and A*STAR through the Industry Alignment Fund - Industry Collaboration Projects Grant.

\clearpage

%

{\small
\bibliographystyle{ieee_fullname}
\bibliography{egbib}
}
\clearpage

\appendix

{\section*{Appendices}\huge }
\section{Demo Video Descriptions}

In the demo video, we show our prediction results and the comparisons between our predictions and that of the baseline method~\cite{gao20192} on FAIR-Play, MUSIC-Stereo and Youtube-ASMR~\ref{suppsec4}, respectively. The video is encoded in H.264 codec.

\section{Separation Training Details}

The task of visually guided sound source separation~\cite{zhao2018sound,gao2019co} aims at separating a mixed audio into independent ones, according to their sound source's visual appearances. We adopt the setting with a mixture of two sources, which is kept the same as Sep-Stereo~\cite{zhou2020sep} for fair comparisons. The backbone audio network $\text{Net}_a$ is shared across stereo and separation learning, following~\cite{zhou2020sep}. 

During training, we follow the Mix-and-Separate pipeline~\cite{zhao2018sound,gao2019co} to mix two mono audios ($s_a$ and $s_b$) of two solo videos as input in the form of STFTs. This can be written as $S_{mix} = S_a + S_b$. The network renders complex masks $\mathcal{M}_a$ and $\mathcal{M}_b$ to predict the STFTs of the audios themselves. We use two APNet~\cite{zhou2020sep} structures for the prediction of complex masks. The loss function can be written as:
\begin{align}
    \mathcal{L}_{sep} = || S_a - \mathcal{M}_a * S_{mix} ||^2_2 + || S_b - \mathcal{M}_b * S_{mix} ||^2_2
\end{align}
This training is done in parallel with stereo, thus the overall loss function is:
\begin{align}
    \mathcal{L} = \mathcal{L}_{stereo} + \lambda_{sep}\mathcal{L}_{sep}.
\end{align}
The weight $\lambda_{sep}$ is empirically set to 1 in our experiments. We observe that it would not affect the results much when $\lambda_{sep} \in [0.5, 1.5]$.

\begin{figure}[t]
    \centering
    \includegraphics[width=0.97\linewidth]{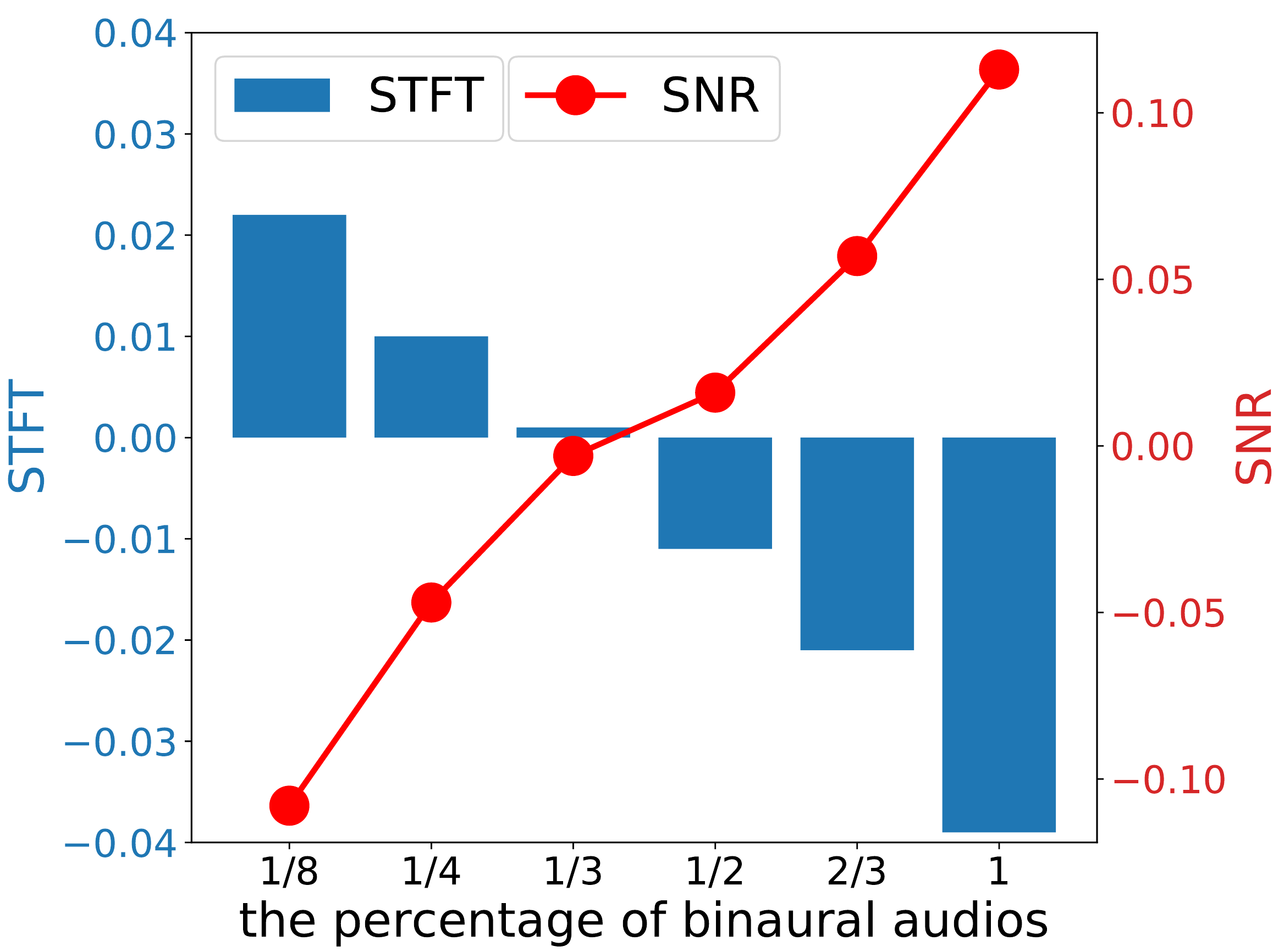}
    \vspace{0pt}
    \caption{The curve of our relative performances \textit{w.r.t} the percentage of binaural audios we use on FAIR-Play. It can be observed that our approach can reach their full performance using only $1/3$ of the ground-truth binaural audios.
}
\vspace{0pt}
\label{fig:amountgt}

\end{figure}

\begin{table*}[t] 
\setlength{\tabcolsep}{12pt}
\begin{center}
\caption{Quantitative results of binaural audio generation on Youtube-ASMR dataset with five evaluation metrics. Note that, our method PseudoBinaural still outperforms Mono-Mono on the non-musical dataset. Owing to the huge data amount and relatively simpler scenarios in Youtube-ASMR dataset, Augment-PseudoBinaural can only achieve a minor improvement over the supervised counterpart~\cite{gao20192}. 
}
\label{table:expASMR}
\begin{tabular}{lccccc}
\toprule
Method & $\text{STFT} $ & $\text{ENV} $ & $\text{Mag} $ & $\mathcal{D}_{phase} $ & $\text{SNR} \uparrow $ \\
\midrule
Mono-Mono  &0.286 &0.070 &0.571 &1.570 &2.111 \\
Mono2Binaural~\cite{gao20192}  &0.198 &0.055 &0.395 &1.396 &3.855 \\
\textbf{PseudoBinaural} (Ours)  &0.252 &0.064 &0.504 &1.517 &2.634 \\
\textbf{Augment-PseudoBinaural}  &\textbf{0.196} &\textbf{0.055} &\textbf{0.394} &\textbf{1.394} & \textbf{3.860} \\

\bottomrule
\end{tabular}
\end{center}
\vspace{0pt}
\end{table*}

\section{Implementation Details}
Following \cite{gao20192}, we use Python package Audiolab to resample the audio at 16kHz, which significantly accelerates the IO process during training. Besides, our follow the same details for our network structure and the training protocol as in Mono2Binaural~\cite{gao20192} and Sep-Stereo~\cite{zhou2020sep}. The spectrograms are of size $257 \times 64 \times 2$, and the visual inputs are $224 \times 448 \times 3$ images. During testing, we find an inappropriate normalization operation in the demo-generating script presented in the public code of \cite{gao20192} and~\cite{zhou2020sep}. Specifically, since the ground-truth binaural audio is unknown in advance, the normalization step should be implemented in the mono audio instead. After fixing this bug, we discover the hop length of sliding window will not affect the inference performance. Hence, the sliding window is set to 0.1s for all the experiments.

While creating the pseudo visual-stereo pairs, we leverage the off-the-shelf human detector and instrument detector of Faster RCNN~\cite{ren2015faster, gao2019co} to crop the visual patches from the videos with mono audios. The center of the cropped patches is placed in an arbitrary position on a blank background. For the generation of pseudo binaurals, the number of speakers in the virtual array is defined as $N = 8$ following the common practice. The speakers are uniformly placed around the frontal part of the head.

As for the activation map in Fig.4, it comes from the feature map of the last convolutional layer in the visual network. For a specific image input, the corresponding feature map is averagely pooled along the channel dimension and normalized to $[0,1]$. And then, we deploy a bilinear upsampling operation on the feature map, making the size align with the original input image. In the end, we set the transparency ratio of upsampled feature map as 0.4, and combine it with the input image to obtain the final activation map.

\section{More Ablation Studies}
\subsection{Ablation Study on the Amount of Binaural Recordings}
Similar to the extensive analysis in~\cite{zhou2020sep}, we provide an ablation study on the amount of binaural recording audios used in the augmentation experiments. Fig~\ref{fig:amountgt} shows the relative performance gains \textit{w.r.t} the percentage of binaural audios used on FAIR-Play. The curve is drawn in a relative setting, where the performance of Mono2Binaural serves as the reference (zero in both metrics). As illustrated in Fig~\ref{fig:amountgt}, Augment-PseudoBinaural can achieve the comparable performance based on only $1/3$ of the ground-truth binaural audios, which further demonstrates the effectiveness of our proposed method.

\begin{table}[t] \footnotesize
\vspace{-2pt}
\setlength{\tabcolsep}{12pt}
\begin{center}
\caption{Results for different mixes of $K$ on FAIR-Play.}
\label{table:ratio}
\begin{tabular}{lcccc}
\toprule
Mix ratio & $ 1:1:1 $ & $4:5:1$ & $4:1:5 $ & $1:5:4 $ \\
\midrule
STFT $\downarrow$  &0.880 &\textbf{0.878} &0.884 &0.885 \\
SNR $\uparrow$  &5.312 &\textbf{5.316} &5.310 &5.306 \\

\bottomrule
\end{tabular}
\end{center}
\vspace{-10pt}
\end{table}

\subsection{Ablation Study on Mix Ratios of $K$}
We conduct an ablation study on the selection of different mix ratios. The portions are the partitions of videos with one, two and three mixed sources used during training. During our implementation, the ratios are selected as 4 : 5 : 1. The ablation results on FAIR-Play are listed in Table~\ref{table:ratio}. It indicates that the influence of different mix ratios is minor.

\section{Experiment on Youtube-ASMR Dataset}
\label{suppsec4}

Youtube-ASMR introduced by Yang \textit{et al}.~\cite{yang2020telling} is a large-scale video dataset collected from YouTube. It consists of approximately 300K 10-second video clips with spatial audio and lasts about 904 hours in total. In an ASMR video, only an individual actor or ``ASMRtist'' is speaking or making different sounds towards a dummy head or the microphone arrays while facing the cameras. Compared to FAIR-Play or MUSIC-Stereo, the binaural scenarios in Youtube-ASMR dataset are relatively simpler since a typical sample just contains one sound source and the visual scene only involves a face in most cases. Similarly, the supervised method Mono2Binaural~\cite{gao20192} and our approach are also implemented on this non-musical dataset. As shown in Tabel~\ref{table:expASMR}, PseudoBinaural can still surpass Mono-Mono on all the metrics, while the augmentation version only improves modestly over the supervised paradigm~\cite{gao20192}.

\section{Limitations}

There are also certain limitations in our work. The first is the domain gap between self-created images and real images. One direction towards solving this problem is through better blending techniques to the background. The second is that we do not specifically model room reverberations, particularly cannot be adapted to any given environment. Domain adaptation techniques for vision might be useful for it. Moreover, our method does rely on the videos that contains only one visual and auditory sound source.

\end{document}